\documentclass[9pt,twocolumn,twoside]{osajnl}
\journal{optica} % Choose journal (ao,jocn,josaa,josab,ol,optica,pr)

\setboolean{shortarticle}{false}
\usepackage{lineno}
\usepackage{subcaption}
\DeclareUnicodeCharacter{2009}{\,} 
\title{Spatio-temporal pulse cleaning in multi-pass cells}

\author[1]{Jaismeen Kaur}
\author[1*]{Louis Daniault}
\author[1]{Zhao Cheng}
\author[1]{Oscar Tourneur}
\author[2]{Olivier Tcherbakoff}
\author[2]{Fabrice Réau}
\author[2]{Jean-François Hergott}
\author[1]{Rodrigo Lopez-Martens}

\affil[1]{Laboratoire d'Optique Appliqu\'ee, Institut Polytechnique de Paris, ENSTA-Paris, Ecole Polytechnique, CNRS, 91120 Palaiseau, France}
\affil[2]{Université Paris-Saclay, CEA, CNRS, LIDYL, 91191, Gif-sur-Yvette, France}

\affil[*]{Corresponding author: louis.daniault@ensta-paris.fr}

\begin{abstract}
We study both numerically and experimentally the use of two third-order nonlinear temporal filtering techniques, namely nonlinear ellipse rotation (NER) and cross-polarized wave (XPW) generation, for spatio-temporal cleaning of mJ energy 30\,fs Titanium:Sapphire laser pulses in a multi-pass cell. In both cases, a contrast enhancement greater than 3 orders of magnitude is observed, together with excellent output pulse quality and record high conversion efficiencies. Careful balancing of nonlinearity and dispersion inside the multi-pass cell helps tune the spectral broadening process and control the post-compressed pulse duration for specific applications. 
\end{abstract}

\setboolean{displaycopyright}{true}

\begin{document}

\maketitle

\section{Introduction}
In the past few years, multi-pass cells (MPC) have emerged as efficient tools for nonlinear spectral broadening and temporal manipulation of ultrashort laser pulses~\cite{MPCtheory_Hanna2017, MPC_Schulte2016}, mainly because of their high throughput~\cite{MPC_Weitenberg2017, MPC_gasFilled_Lavenu2018, MPC_gasFilled_Kaumanns2018, MPC_HighCompressionFactor_Balla2020, HighPower_FewCycle_Muller2021}, power scalability~\cite{MPC_highestPeakPower_Kaumanns2021, Spatio-spectral_Daher2020, MPC_KWpower_Grebing2020}, high compression factors~\cite{MPC_HighCompressionFactor_Balla2020}, spatio-spectral beam homogeneity~\cite{MPC_temporalQuality_Escoto2022}, and strong potential for nonlinear spatio-temporal pulse shaping applications~\cite{MPC_Weitenberg2017, MPC_Serrodyne_Balla2022}. The attractiveness of MPCs lies in the fact that extended nonlinear propagation occurs over a relatively small footprint and that large B-integrals can be accumulated in small increments for every pass with minimal spatio-temporal couplings, provided the input beam size is carefully matched to the cell eigenmode and the beam size on the end-mirrors remains fairly constant for every pass \cite{ModeMatching_Hanna2021}. MPCs also provide a large number of degrees of freedom in terms of choice of nonlinear medium as they are compatible with bulk material \cite{MPC_Schulte2016}, gases \cite{MPC_gassFilled_Ueffing2018, MPC_gasFilled_Lavenu2018, MPC_gasFilled_Kaumanns2018} or even hybrid geometries \cite{HybridMPC_Seidel2022}. 

Recently, MPC-based post-compression was extended from Ytterbium(Yb)-based laser systems to Titanium:Sapphire (Ti:Sa) lasers, with compressibility down to the few-cycle regime in a single stage~\cite{SN3_MPC_Daniault2021, MPCwithTiSa_Rueda2021}. When used to reach ultra-high focused intensities in the frame of light-matter interaction experiments, most Ti:Sa laser systems rely on some form of nonlinear temporal filtering to suppress amplified spontaneous emission (ASE) and parasitic pulses surrounding the pulse peak. Such contrast enhancement techniques include saturable absorbers~\cite{SaturableAbsorbers_ITATANI1998, Saturableabsorber_Fourmaux2011}, optical parametric amplifiers~\cite{OPA_DUBIETIS1992, OPA_Yoshida2003, OPA_Dorrer2007}, second harmonic generation~\cite{SHG_Aoyama2001}, plasma mirrors~\cite{plasmaMirrors_Kapteyn1991, DoublePM_Levy2007}, spectrally filtered self-phase modulation (SPM)~\cite{FilteredSPM_Buldt2017}, nonlinear ellipse rotation (NER)~\cite{NERat1um_Tapie1992, NER_Sala1978, NERinHCF_Homoelle2002, NER_Kalashnikov2004, NER_LOA_Jullien2004, NER_Chun_Mei2008} and the widely used cross-polarized wave generation (XPW)~\cite{ContrastEnhancementXPW_Julien2005}. 

In this article, we benchmark both NER and XPW techniques in an MPC architecture for simultaneous post-compression and spatio-temporal cleaning of mJ-energy 30\,fs pulses from a Ti:Sa laser, with compressibility down to few-cycle duration and record efficiency. We also rely on comprehensive (3+1)D numerical simulation of the nonlinear process, which accurately reproduces the measured data, to pinpoint the role played by the different experimental parameters and to find optimized, application-specific configurations for both techniques. 

\section{Nonlinear ellipse rotation in multipass cells}

In NER, temporal cleaning is achieved thanks to the intensity-dependent rotation of the polarization ellipse of an ultrashort pulse undergoing nonlinear propagation \cite{NLO_Boyd2008}. Polarizing optics can be used to discriminate the high-intensity pulse peak that experiences nonlinear rotation against the unwanted, unrotated low-intensity ASE pedestal and the parasitic side pulses. NER was tested early on for contrast enhancement and spatial mode cleaning in air~\cite{NER_Kalashnikov2004, NER_LOA_Jullien2004} and gas-filled hollow fibers~\cite{NERinHCF_Homoelle2002}, the latter having been later shown to enable post-compression down to few-cycle duration with internal conversion efficiencies in the range of 40-55\%~\cite{NERinHCF_Khodakovskiy2019, NERinHCF_Smijesh2019}. Very recently, NER in a gas-filled MPC was first explored through simulations~\cite{NERinMPC_Pajer2021} and then experimentally with Yb lasers both in a gas~\cite{NERinMPC_Pfaff2022} and in a multi-plate arrangement~\cite{NER_Seidel2022}. In the following section, we describe how NER implemented in a gas-filled MPC can be used as a direct method for generating few-cycle pulses with high temporal fidelity using a Ti:Sa laser. 

\subsection{Experimental setup for NER}

\begin{figure}[htbp]
\centering
\includegraphics[width=\linewidth]{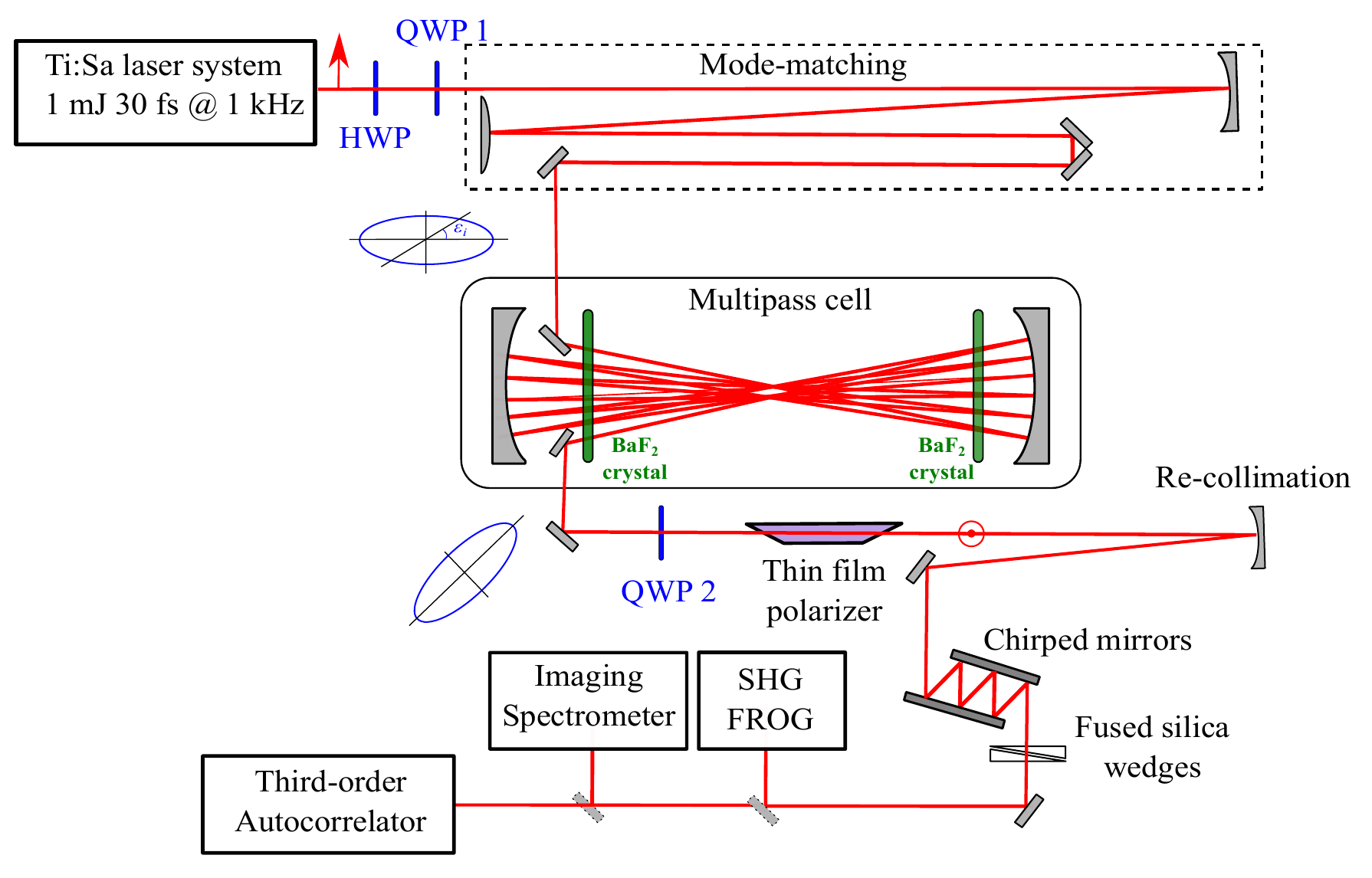}
\caption{Experimental setup layout for both NER- and XPW-MPC. Components marked in dark blue and green are used for performing NER and XPW, respectively. For clarity, only a few passes in the cell are represented. HWP: Half-Wave Plate, QWP: Quarter-Wave Plate}
\label{fig:expt-setup}
\end{figure}

A schematic layout of the general experimental setup is shown in fig. \ref{fig:expt-setup}. A Ti:Sa chirped pulse amplifier system (Femtopower Pro-HE) delivers mJ-energy 30\,fs pulses at 1\,kHz repetition rate and features an acousto-optic programmable dispersion filter (Dazzler, Fastlite) that enables precise control over the output pulse spectral phase. A pair of focusing/diverging optics is used to fulfill the mode-matching conditions for stable MPC operation. A pair of folding mirrors mounted on a translation stage after the telescope is used to tune the position of the beam-waist inside the cell. The quasi-concentric MPC consists of two 50.8\,mm diameter enhanced-silver coated mirrors with a radius of curvature of 1.5\,m, separated by $\sim 3$\,m. A small off-axis plane injection mirror couples the beam into the cavity with each beam pass forming a circular pattern on the mirror. After the required number of round-trips, the beam is picked off by another small plane off-axis mirror located on the same side as the injection. The 50.8\,mm cavity mirror size ensures sufficient distance between the adjacent beam spots on the mirrors and minimizes beam steering losses. The number of passes in the cavity can be varied in even numbers by simply moving the pick-off mirror and can be set to a maximum of 18. The MPC can therefore support up to 54\,m of propagation folded within an effective 3\,m footprint. 

The MPC is filled with Argon at a controlled pressure. Nonlinear refraction in Argon is taken into account in the simulations to find the appropriate mode-matching conditions and maintain an overall constant beam size on the cavity end-mirrors and at the center of the cavity, which also avoids damaging the optics and minimizes ionization~\cite{PropModelMPC_Hanna2020}. The beam waist is set to $450~\mu\mathrm{m}$, corresponding to a 2.9\,mm beam diameter on the end-mirrors. The number of passes is set to 12, corresponding to a total propagation length of 36\,m. 

Elliptical polarization is achieved by inserting a broadband quarter-wave plate (QWP) into the collimated beam before entering the mode-matching telescope. A half-wave plate (HWP) is placed before it to tune the input polarization direction and hence the ellipticity $\epsilon_i$ inside the MPC from 0 (linear) to 1 (circular). A second crossed broadband QWP is used to retrieve linear polarization at the output of the MPC. A low-dispersion broadband thin film polarizer (TFP) (Femtolasers GmbH) with an extinction ratio of $5 \times 10^{-3}$ filters out the pulse peak rotated by NER at 90° with respect to the input polarization, while rejecting the low-intensity unrotated background. The temporally filtered pulses are then post-compressed using a set of chirped mirrors (up to $-~450~\mathrm{fs^2}$, PC42, Ultrafast Innovations GmbH) and a pair of adjustable thin fused silica wedges. A range of diagnostic tools are used to characterized the NER-filtered pulses: An SHG-FROG (FemtoEasy) characterizes the pulses spectrally and temporally, a high dynamic range third-order cross-correlator (TUNDRA, Ultrafast Innovations GmbH) is used to detect the ASE pedestal and parasitic pulses over a time window of 200\,ps around the pulse peak, and an imaging spectrometer (MISS, FemtoEasy) is used to assess the output spatio-spectral beam quality. 

\subsection{Maximizing NER efficiency}

Internal NER efficiency is defined as the ratio between the power measured after and before the output polarizer. Maximum NER efficiency is therefore achieved when the polarization ellipse rotates by exactly 90°, such that transmission through the TFP is maximized. However, the NER angle is time-dependant along the time-varying pulse profile, which leads to a time-dependent transmission through the TFP. Moreover, as the pulses simultaneously experience spectral broadening and chirping through self-phase modulation (SPM), the NER angle also becomes wavelength-dependant, implying that the broadened spectrum is not uniformly transmitted through the polarizer. All these effects combined limit both the energy and the spectral throughput and thus drastically affect post-compression performance.

\begin{figure}[htbp]
\centering
 \includegraphics[width=\linewidth]{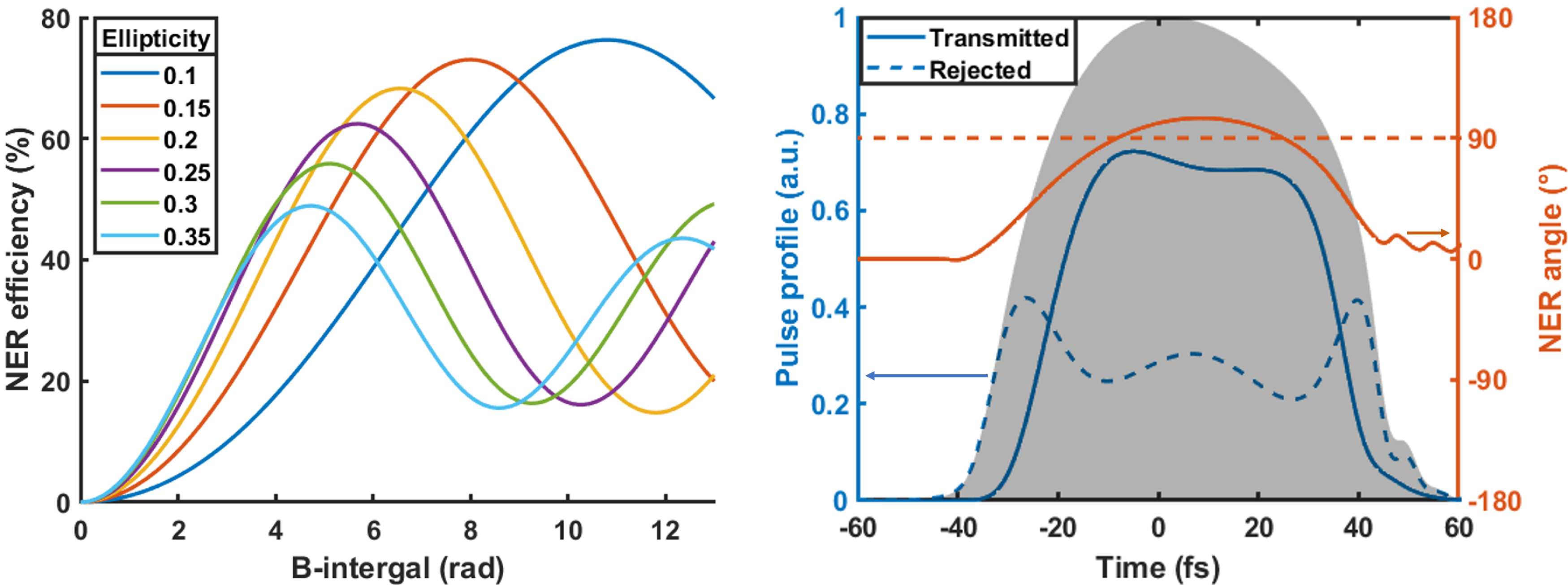}
 \caption{Left: Simulated NER efficiency vs. B-integral for various pulse ellipticities. Right: Temporal profile of NER pulse (grayed), transmitted pulse (solid blue) and rejected (dashed blue) pulse by the TFP; time-dependant NER angle (solid red), all simulated for $\epsilon_i = 0.25$ at maximum efficiency ($B\simeq 6\:$ rad).}
\label{fig:NER_Simus}
\end{figure}

Fig. \ref{fig:NER_Simus} (left panel) shows the evolution of internal NER efficiency versus B-integral for different ellipticities, obtained from numerical simulations for a plane-wave 1\,mJ, 30\,fs pulse propagating in Argon, including SPM, cross-phase modulation (XPM), self-steepening and gas dispersion. One clearly notices that lower ellipticities lead to a reduced spread of NER angles and, when combined with high B-integrals, yield higher throughput. However, our previous work on direct post-compression of 1\,mJ 30\,fs TiSa pulses in the same MPC configuration showed that the B-integral cannot be pushed much beyond 6 rad, corresponding to an Ar gas pressure around 400\,mbar, where excessive self-steepening effects lead to poor pulse compressibility~\cite{SN3_MPC_Daniault2021}. In the case of NER, the ellipticity should therefore not be lower than 0.15 and the maximum NER efficiency should not exceed 75\%. Fig. \ref{fig:NER_Simus}(right panel) shows the simulated time-dependent NER angle during the rise and fall of the pulse along with the transmitted and rejected pulse profiles for $\epsilon_i = 0.25$ at maximum efficiency ($B\simeq 6\:$ rad). The maximum NER angle is above 90°, such that the transmission averaged over the whole pulse profile is maximized. 

\begin{figure}[h]
\centering
\includegraphics[width=\linewidth]{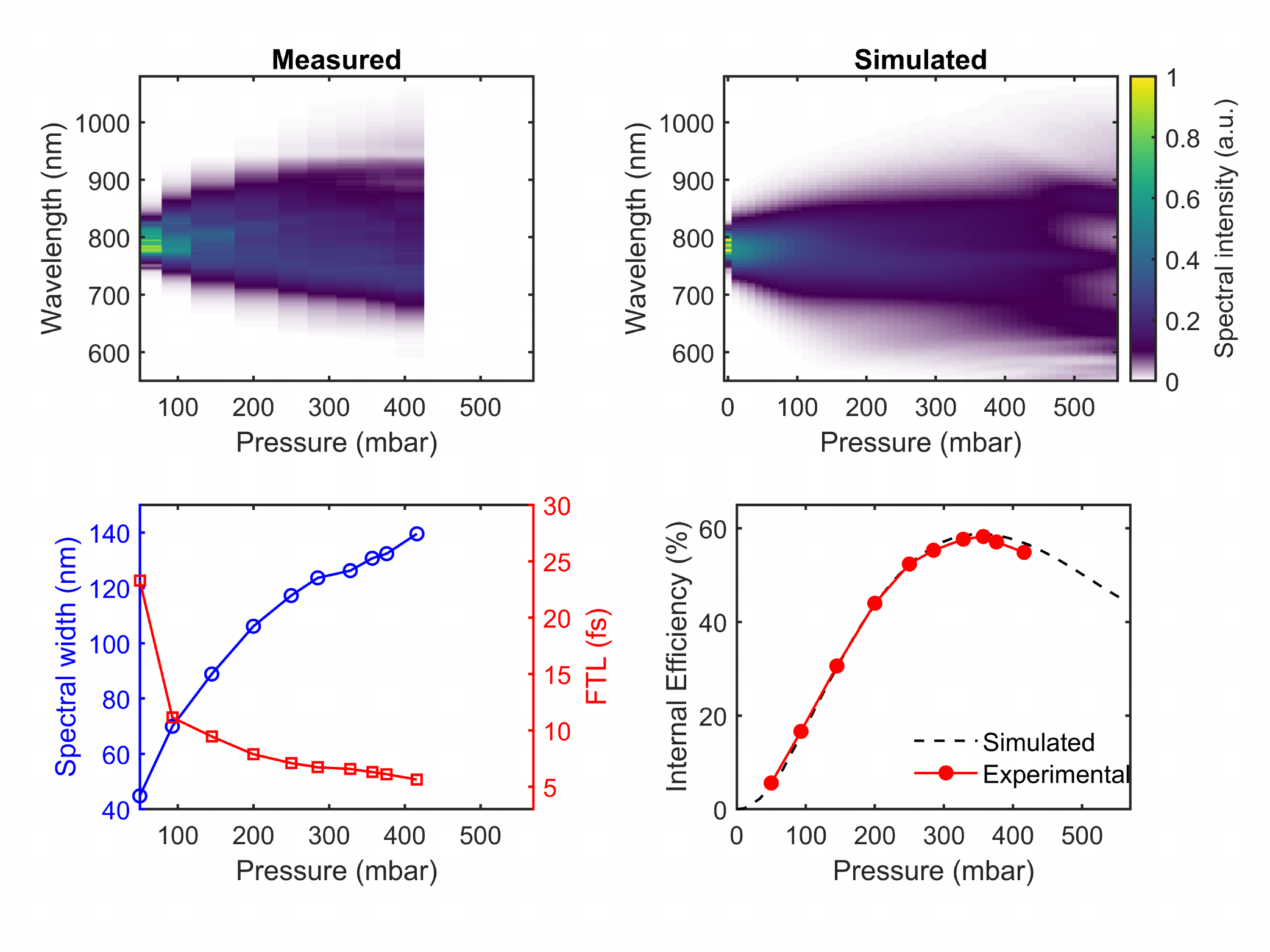}
\caption{Top: evolution of the measured (left) and the simulated (right) transmitted NER spectrum with Ar gas pressure, for $\epsilon_i=0.25$. Bottom: variation of the measured spectral width and Fourier-transform-limited (FTL) duration (left) and corresponding experimental NER conversion efficiency with Ar pressure compared to simulated values (right).}
\label{fig:NERevolution}
\end{figure}

We performed the experiment using $\epsilon_i = 0.25$ and measured the internal NER efficiency along with the output pulse spectrum for increasing Ar pressures up to 420\,mbar, where pulse break-up starts. Fig. \ref{fig:NERevolution} compares the measured data with the results obtained from (3+1)D simulations now including temporal, radial, and polarization dimensions and using measured device losses and the experimental laser pulse spectrum as input. Simulations show excellent agreement with measurements. Experimentally, the spectral bandwidth increases fast at first, then starts flattening out before increasing again around 420\,mbar. The effects of pulse break-up can be seen in simulations at higher pressures, with the sudden appearance of wings and deep modulations in the broadened spectrum. The agreement between experiment and simulations is particularly good for the internal NER efficiency versus pressure, which reaches a maximum of 58\% at an optimum Ar pressure of $\simeq$~350 mbar and then rolls off because of the peak NER angle exceeding 90°. We now can tune the ellipticity to a lower value, such that the maximum NER efficiency occurs just before pulse break-up around 420\,mbar. This should lead to both a higher throughput and a broader spectrum, and therefore yield shorter compressed output pulses.

\subsection{Experimental results at optimum ellipticity}

\begin{figure}[h!]
\centering
\includegraphics[width=\linewidth]{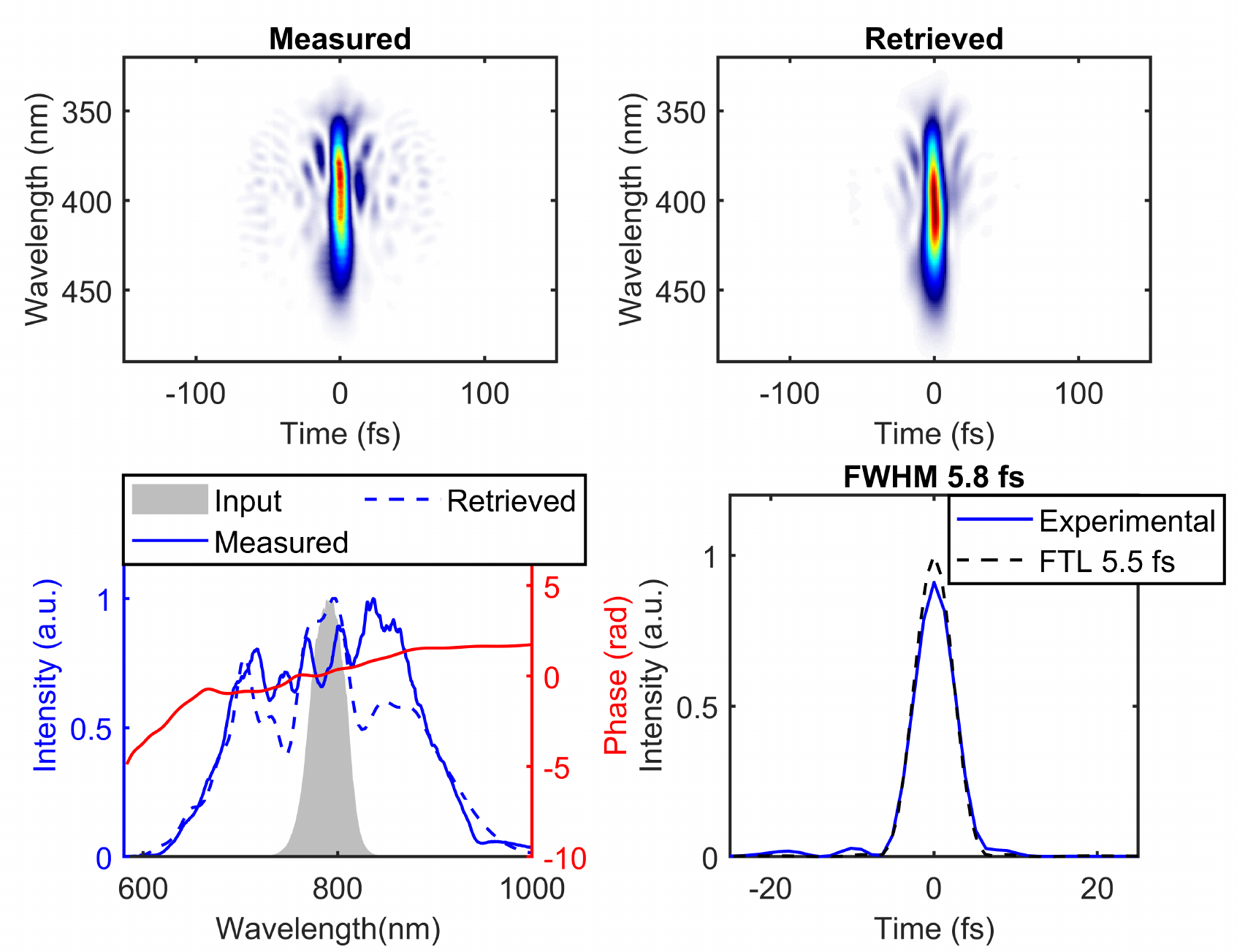}
\caption{Top: measured (left) and retrieved (right) SHG-FROG traces. Bottom: input and NER spectra (left) and the corresponding temporal profile (right) for $\epsilon_i = 0.18$.}
\label{fig:NER-FROG}
\end{figure}

By setting the Ar pressure to 420\,mbar and the pulse ellipticity to $\epsilon_i=0.18$, the internal NER efficiency increases from 58\% to 69\%. The output polarizer is rotated accordingly to preserve the extinction ratio and the output energy drops to 0.49\,mJ, yielding a global NER efficiency of 49\%, including losses in the MPC. In this configuration, we obtain, nearly-FTL 5.8 fs pulses (2.2 optical cycles at $\lambda=800\ $nm) as shown in fig. \ref{fig:NER-FROG}. The reconstructed 5.8\,fs pulse profile (solid blue curve) is very close to the Fourier-transform-limited profile (FTL, dotted black curve) exhibiting near-perfect compression with a very low-intensity pedestal structure, limited only by residual phase and spectral modulations introduced by the paired double-angle chirped mirror compressor. The compressor efficiency is measured to be 87\%, leading to an overall post-compression efficiency of 43\%.

\begin{figure}[h]
\centering
\includegraphics[width=0.8\linewidth]{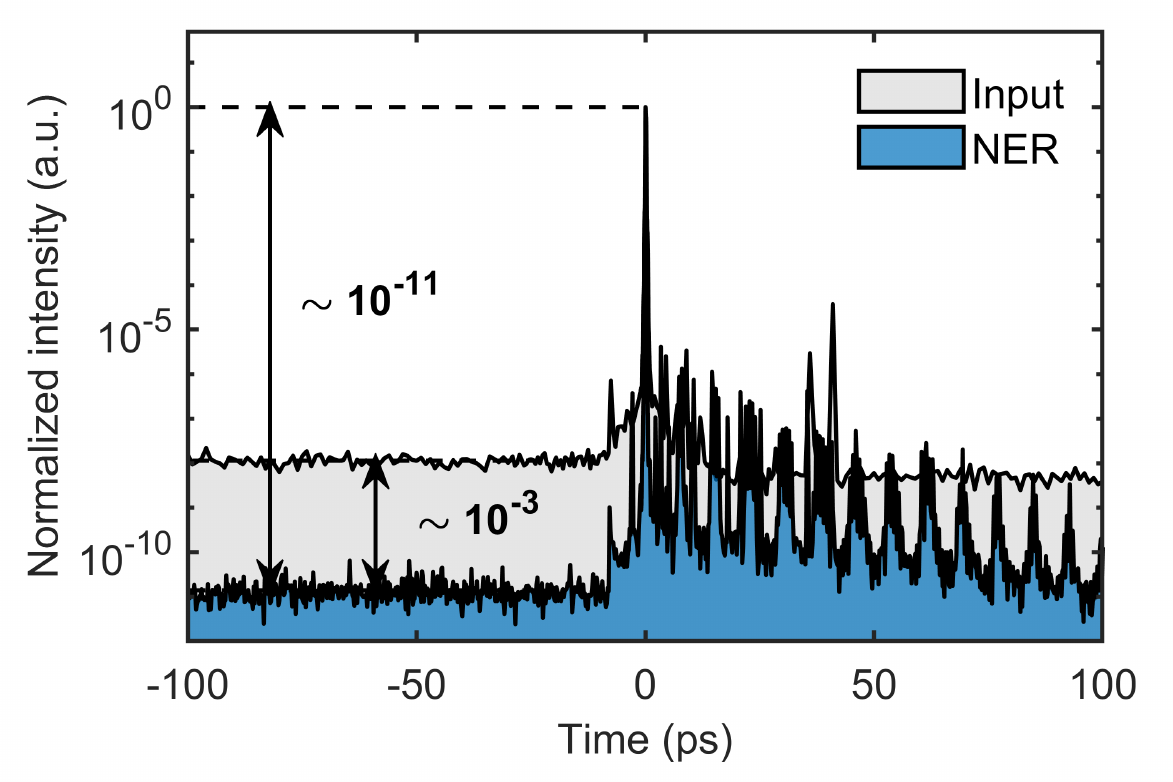}
\caption{Temporal contrast enhancement between input and NER ($\epsilon_i=0.18$) pulses measured using a high dynamic range third-order autocorrelator.}
\label{fig:NER_contrast}
\end{figure}

\begin{figure}[h]
\centering
\includegraphics[width=\linewidth]{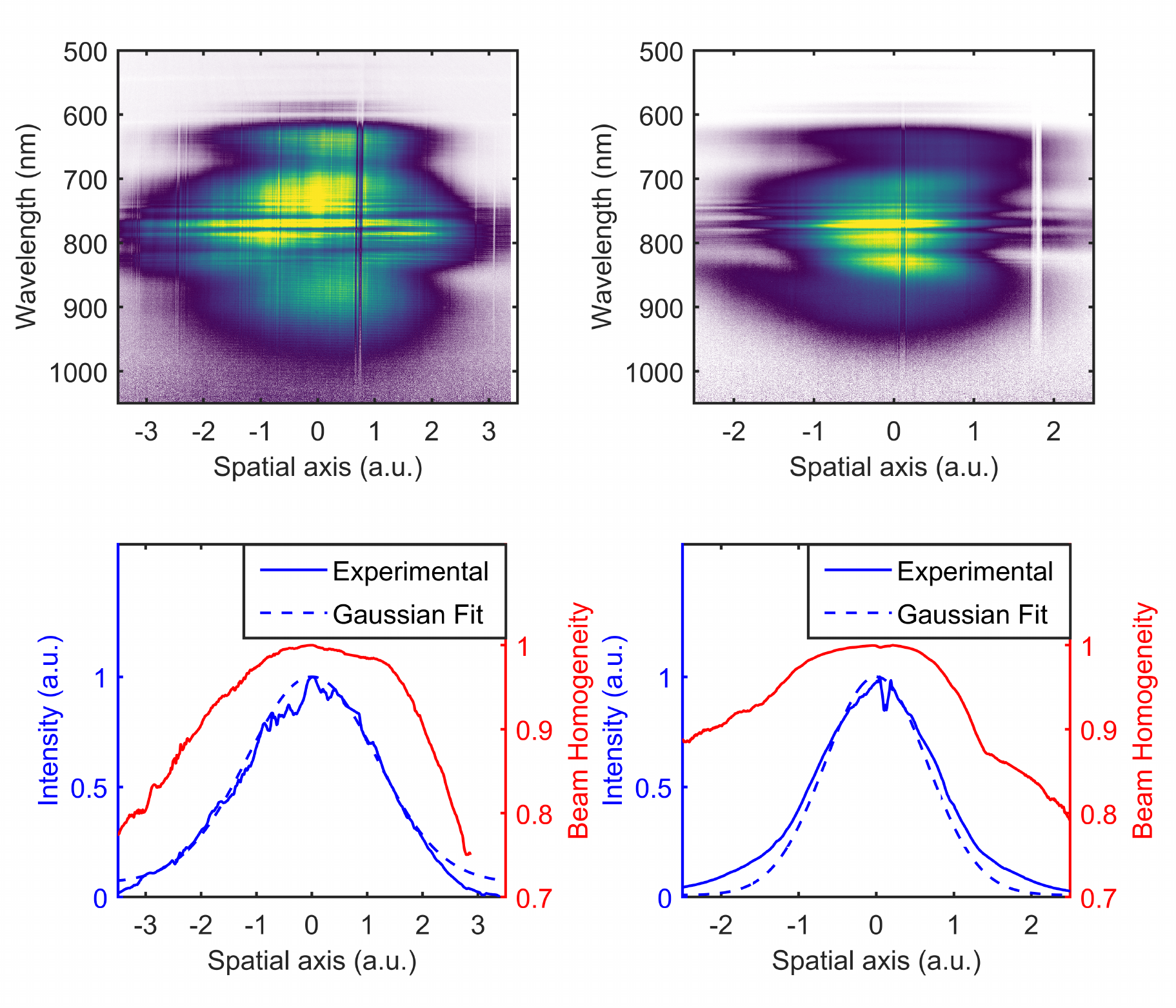}
\caption{(Top) spectrally-resolved beam profile in the horizontal (left) and vertical (right) dimensions and (bottom) output beam profile in arbitrary units in the horizontal (left) and vertical dimensions (right) along with their spectral homogeneity, measured for $\epsilon_i=0.18$. }
\label{fig:NER_spatio-specrtral}
\end{figure}

Fig. \ref{fig:NER_contrast} compares the long-range temporal profiles of input and output pulses measured by the TUNDRA device. The contrast enhancement obtained via NER is at least 3 orders of magnitude, with ASE levels dropping down to $1:10^{-11}$ a few ps prior to the pulse peak, and is limited by the extinction ratio of the TFP. The pre-pulse seen at -7.5\,ps in the traces for both the input and NER pulses is an inherent artifact of the TUNDRA. The train of post-pulses visible in the NER trace originates from internal reflections within the TFP itself and are not observed when a Glan polarizer is used for extinction. However, the high thickness of a Glan polarizer leads to excessive dispersion and nonlinear effects distorting the output pulse and beam. Finally, the spectrally-resolved beam profile in the horizontal and vertical dimensions and the corresponding V-parameter measured for $\epsilon_i=0.18$ are shown in fig. \ref{fig:NER_spatio-specrtral}. In both dimensions, the output beam exhibits good spectral homogeneity, $>98$\% at FWHM and above $>85$\% at $1/e^2$.

These results must be compared to direct post-compression experiments~\cite{SN3_MPC_Daniault2021}. Under the exact same experimental conditions, albeit with only 16 passes in the MPC, we measured 5.3\,fs post-compressed pulses with 66\% overall efficiency, chirped mirror compressor included. The 43\% overall transmission measured in the case of NER amply justifies its implementation in an MPC post-compressor as it enables the direct generation of high-contrast few-cycle pulses with moderately higher losses, little compromise on the output pulse duration and very low added complexity.

\section{Cross-polarized wave generation in multi-pass cells}

XPW generation is a well-established and perhaps the most widely used technique for temporal contrast enhancement in high-energy Ti:Sa lasers. It is a degenerate four-wave mixing process governed by the anisotropy of the real part of the third-order nonlinearity tensor $\chi^{(3)}$. Inside the nonlinear medium, a new orthogonally polarized wave is generated at the same central frequency as the incident wave. Conventionally, its implementation is quite straightforward: a linearly polarized laser pulse is focused into a crystal placed between crossed polarizers. Due to the cubic intensity dependence of the process in space and time, efficient conversion occurs only at high intensities, making it easy to filter out the cleaned XPW pulse from the lower-intensity ASE pedestal and parasitic pulses. The XPW conversion efficiency depends on the intensity of the pulse incident on the crystal, the thickness of the crystal, the input spatio-temporal pulse quality and its spectral phase~\cite{ SpectralPhase_XPW_Jullien2007, SpectralPhase_XPW_Canova2008}. The incident intensity on the XPW crystal is limited by the threshold of white light generation in the crystal (e.g. $\sim 10^{12}~\mathrm{W/cm^2}$ for BaF$_2$ crystals). Using thicker crystals to achieve higher conversion leads to unwanted nonlinear third-order processes that tend to compete with XPW generation, making the XPW beam properties more sensitive to spatial-temporal couplings. The input intensity is also limited by damage due to self-focusing inside the crystal, which tends to reduce its lifetime. So far, the highest demonstrated global conversion efficiency has been limited to 10-20\% using a double thin-crystal configuration~\cite{EfficientXPW_Julien2006} and, for mJ energy pulses, some form of spatial filtering or shaping is needed to ensure a smoother or more homogeneous incident spatial beam profile on the crystals~\cite{MoreEfficientXPW_Julien2008, EfficientXPW_Ramirez2011}. In this work, we tested the implementation of XPW in the MPC, since the nonlinearity inside an MPC is acquired in small increments and spatially redistributed across the beam for every pass.

\subsection{Experimental setup for XPW}

The setup for testing XPW in the MPC is depicted in fig. \ref{fig:expt-setup}. Here, no QWPs are needed since the linear polarization direction of the XPW signal can simply be set by the HWP at the MPC input. The chamber is operated under vacuum and two anti-reflection-coated, 600\,µm thick, holographic cut BaF$_2$ crystals aligned with same orientation are placed symmetrically with respect to the center of the MPC. This configuration helps to mitigate spatial nonlinear effects and ensures spatio-spectral beam homogeneity. The distance of the crystals from the waist enables the tuning of the nonlinearity for every pass and is set to about 1\,m, placing the crystals approximately 50\,cm away from the end-mirrors. The chirped mirrors of the output compressor are changed to accommodate the narrower spectral bandwidth typically produced by the XPW process, and to introduce higher dispersion in order to compensate for the total amount of material traversed by the pulses in the BaF$_2$ plates. The input pulse parameters, output polarizer and pulse diagnostics all remain the same as for NER.

In the case of XPW, mode matching into the MPC becomes much more complicated. First, for a Gaussian input pulse in space and time, both spatial and temporal pulse profiles of the XPW wave are shortened by a factor of $\sqrt{3}$ because of the cubic nonlinearity. Therefore, both input and XPW beams do not share the same beam matching conditions and their respective caustics cannot be stabilized together. Moreover, the input pulse peak power is about $5\times 10^{3}$ times higher than the critical power for self-focusing in BaF$_2$, under our experimental conditions. Although the BaF$_2$ plates are thin enough to prevent catastrophic beam collapse, they have to be considered as thin nonlinear lenses that can disturb mode matching and overall propagation in the MPC. Moreover, the repeated energy transfer from the input pulse to the XPW wave leads to changes in the pulse peak power, such that the Kerr lensing experienced through the BaF$_2$ plates is different for both pulses and for every pass, especially for such high nonlinearities. Material dispersion and reflection losses from the plate surfaces further exacerbate this behavior.

\begin{figure}[h]
\centering
\includegraphics[width=\linewidth]{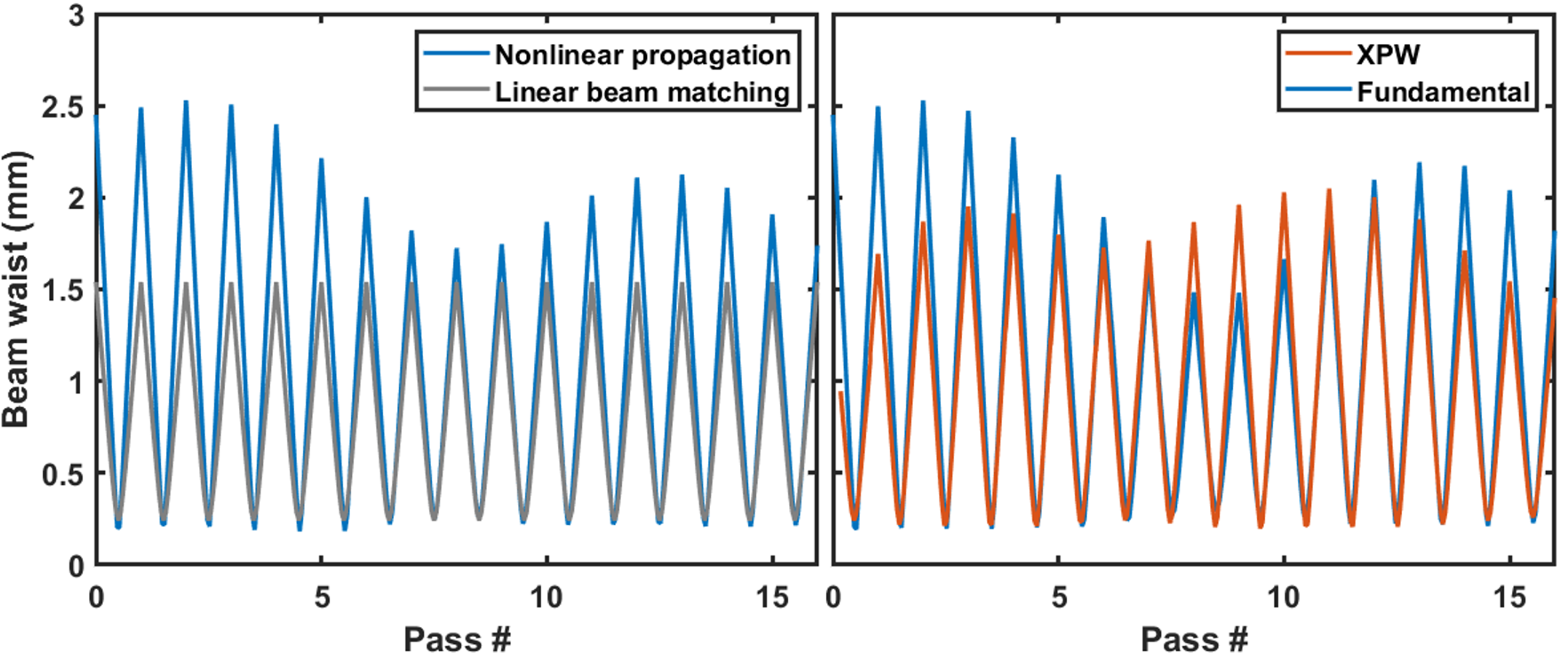}
\caption{Left: Beam caustic of the fundamental wave alone including Kerr Lensing in the BaF$_2$ plates (solid blue) compared to linear mode matching in the MPC (dashed red). Right: XPW and fundamental beam caustics.}
\label{fig:XPW_Caustics}
\end{figure}

Numerical simulations were performed to determine the best beam matching when including Kerr lensing from the BaF$_2$ plates. First, it was run for the input fundamental wave alone, while inhibiting XPW generation. Fig. \ref{fig:XPW_Caustics} shows that the beam caustic of the fundamental wave in the MPC, which is stable for the first few passes and then becomes strongly modulated, illustrating the impact of dispersion on the caustic stability. However, the beam size on the MPC end-mirrors is always larger than that of the linearly matched beam, which excludes any risk of optical damage by design. Excessive ionization at the beam waist, which can occur if the caustic is unstable, is not a concern here as the MPC is operated under vacuum. Second, XPW generation was enabled in the simulation by choosing the proper polarization that maximises efficiency, as detailed in the next section. The XPW caustic is plotted in fig.\ref{fig:XPW_Caustics}. It can be seen that the newly created XPW beam in the first BaF$_2$ plate pass is smaller than the fundamental. As expected, its caustic is highly modulated throughout the MPC, but here again the beam sizes on the mirrors are always larger than in the linear regime. The fundamental beam is more modulated in the presence of XPW generation due to the energy transfer, but the minimum beam size on the mirrors still remains close to that for linear beam matching, thus avoiding optical damages. These simulations show that Kerr lensing in the BaF$_2$ plates drastically disturbs beam propagation and caustic stability throughout the MPC, but also leads to larger beam sizes on the mirrors, such that the beam fluence systematically remains below the damage threshold of the end-mirrors.

\begin{figure}[h]
\centering
\includegraphics[width=0.7\linewidth]{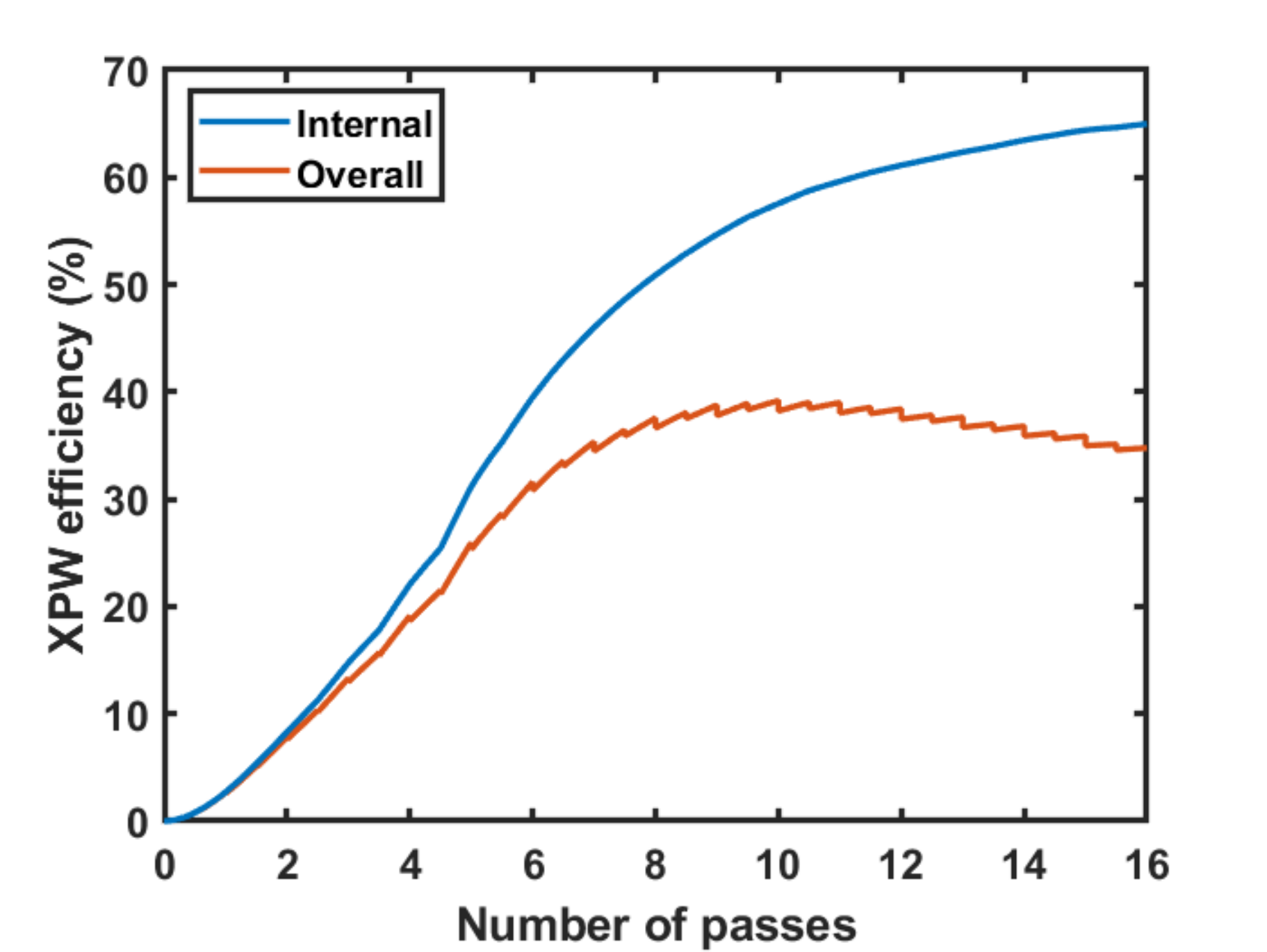}
\caption{Left: Internal (red) and overall (blue) XPW generation efficiency as a function of number of passes in the MPC.}
\label{fig:XPW_efficiency}
\end{figure}

\begin{figure}[h]
\centering
\includegraphics[width=\linewidth]{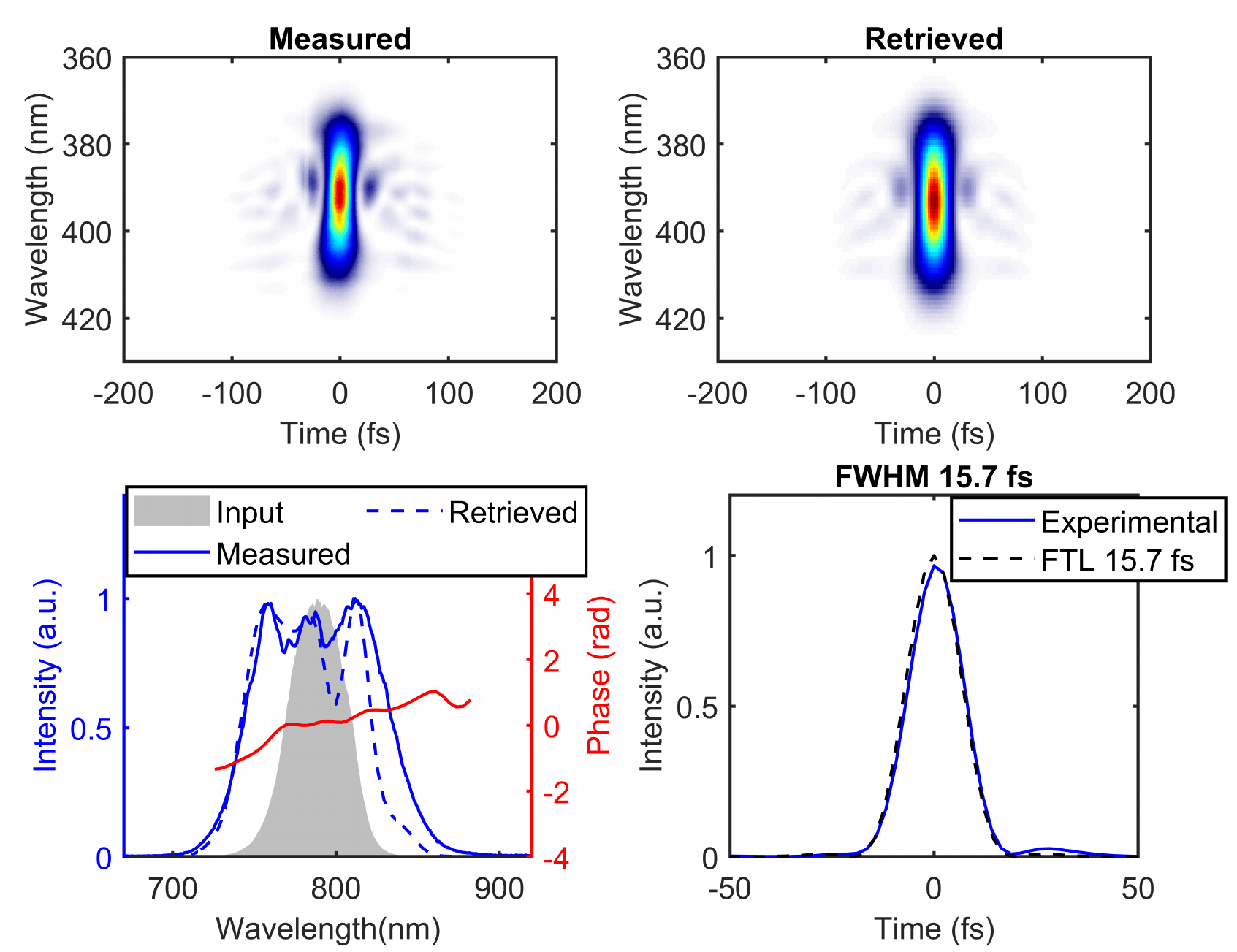}
\caption{Top: measured (left) and retrieved (right) SHG-FROG traces. Bottom: input and XPW spectra (left) and the corresponding compressed XPW temporal profile (right), for FTL pulses injected into the MPC.}
\label{fig:FROG-XPW-FTLcase}
\end{figure}

\subsection{Experimental results of XPW generation}

Fig. \ref{fig:XPW_efficiency} shows the XPW efficiency simulated as a function of the number of passes in the MPC, including reflection losses introduced by the enhanced-silver-coated mirrors ($\sim 1\%$ per bounce) and the BaF$_2$ crystals ($\sim 3\%$ per MPC pass). The input polarization is set to 64.5° with respect to the optical axes of the BaF$_2$ crystals, which maximizes XPW conversion efficiency per pass~\cite{EnergyscalingXPW_Canova2009}. For 16 passes, the total internal XPW efficiency reaches up to 65\%, an unprecedented value which cannot be obtained for XPW in free space. First, the beam size on the crystals is well controlled thanks to the MPC geometry and enables good conversion efficiency per pass, as opposed to a free-space two-plate arrangement, where the beam constantly diverges. Second, the MPC geometry mitigates nonlinear spatial effects, enabling higher efficiencies without distorting the beam profile. However, such a high number of passes through the BaF$_2$ plates implies higher reflection losses, such that the overall efficiency, calculated as the output XPW power over the input power, rapidly drops. In practice, the number of passes should be limited to 10 passes to maximize the overall efficiency to about 35\%, which is still higher than the current state-of-the-art. The beam size is set to $330~\mu$m at the waist and 4\,mm on the end-mirrors in order to fulfill the stability conditions of the MPC. The input pulses are p-polarized and the output TFP is oriented to select the XPW pulses along the orthogonal polarization direction. The XPW pulses are then compressed outside the MPC using a set of chirped mirrors (total dispersion $\simeq -750~fs^2$, HD58 from Ultrafast Innovations) and a pair of adjustable thin fused silica wedges. By compensating dispersion (entrance window, propagation in air) with the Dazzler in the laser chain, we can inject nearly-FTL 30\,fs pulses into the MPC, and obtain $295~\mu$J total XPW pulse energy, corresponding to 57\% internal and 28\% global conversion efficiency, respectively, while taking into account all the losses in the MPC.  

\begin{figure}[h]
\centering
\includegraphics[width=\linewidth]{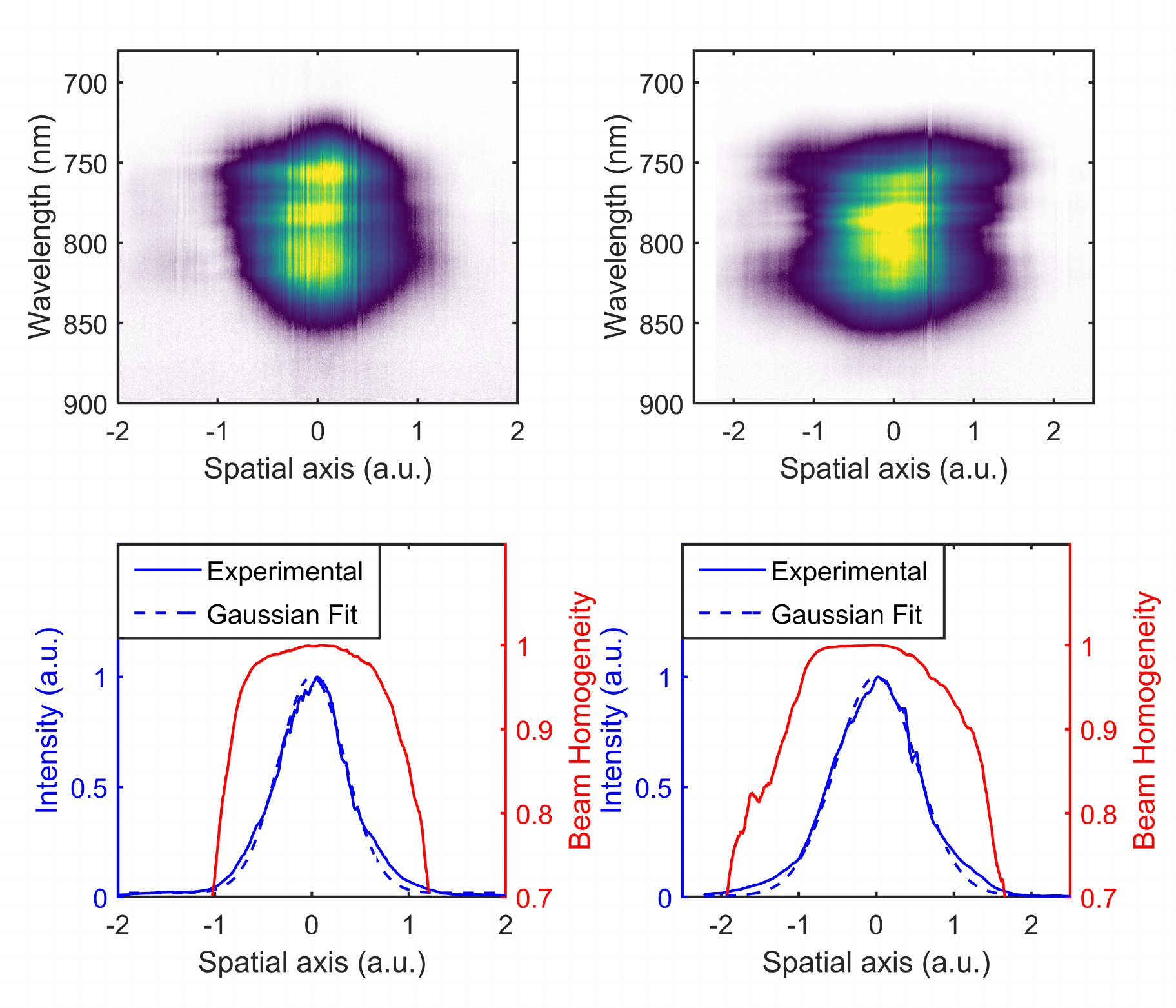}
\caption{Top: measured spatio-spectra traces in horizontal (left) and vertical (right) dimensions; bottom: output beam profile in arbitrary units in horizontal (left) and vertical dimensions (right) along with their spectral homogeneity.}
\label{fig:homogeneity-XPW}
\end{figure}

\begin{figure}[h]
\centering
\includegraphics[width=0.8\linewidth]{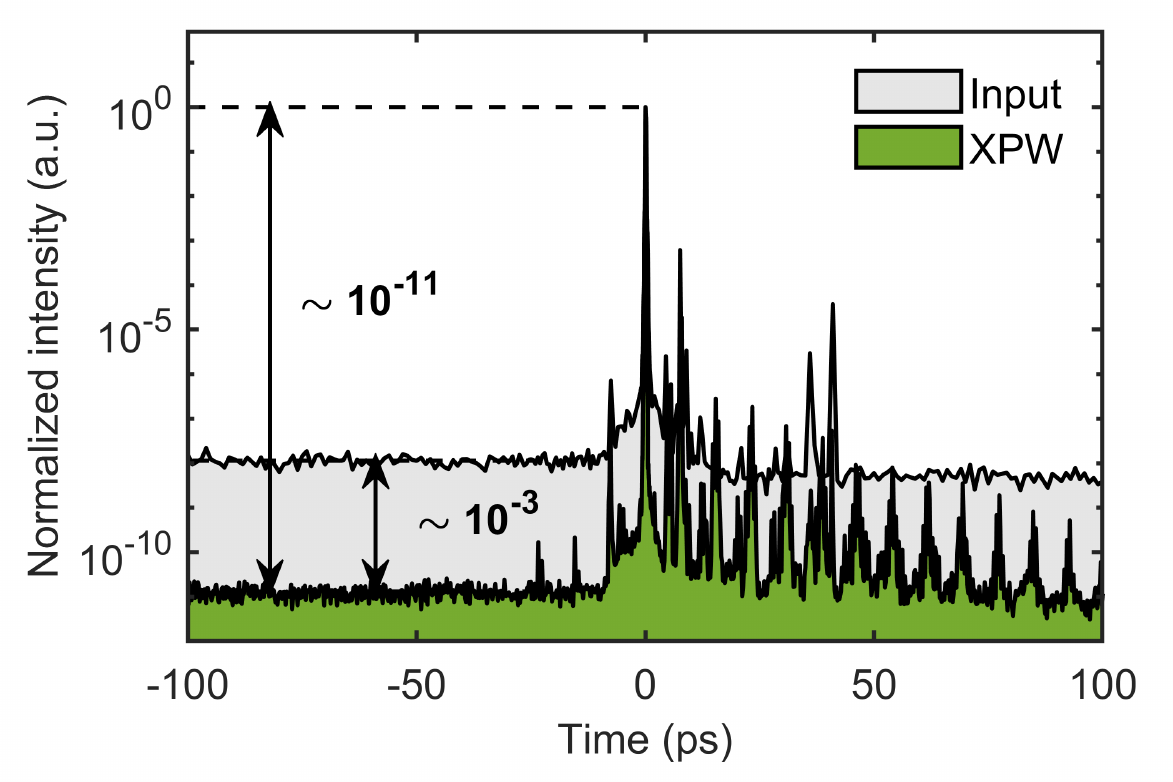}
\caption{Temporal contrast enhancement between fundamental and XPW pulses.}
\label{fig:contrast-XPW}
\end{figure}

\begin{figure}[h]
\centering
\includegraphics[width=\linewidth]{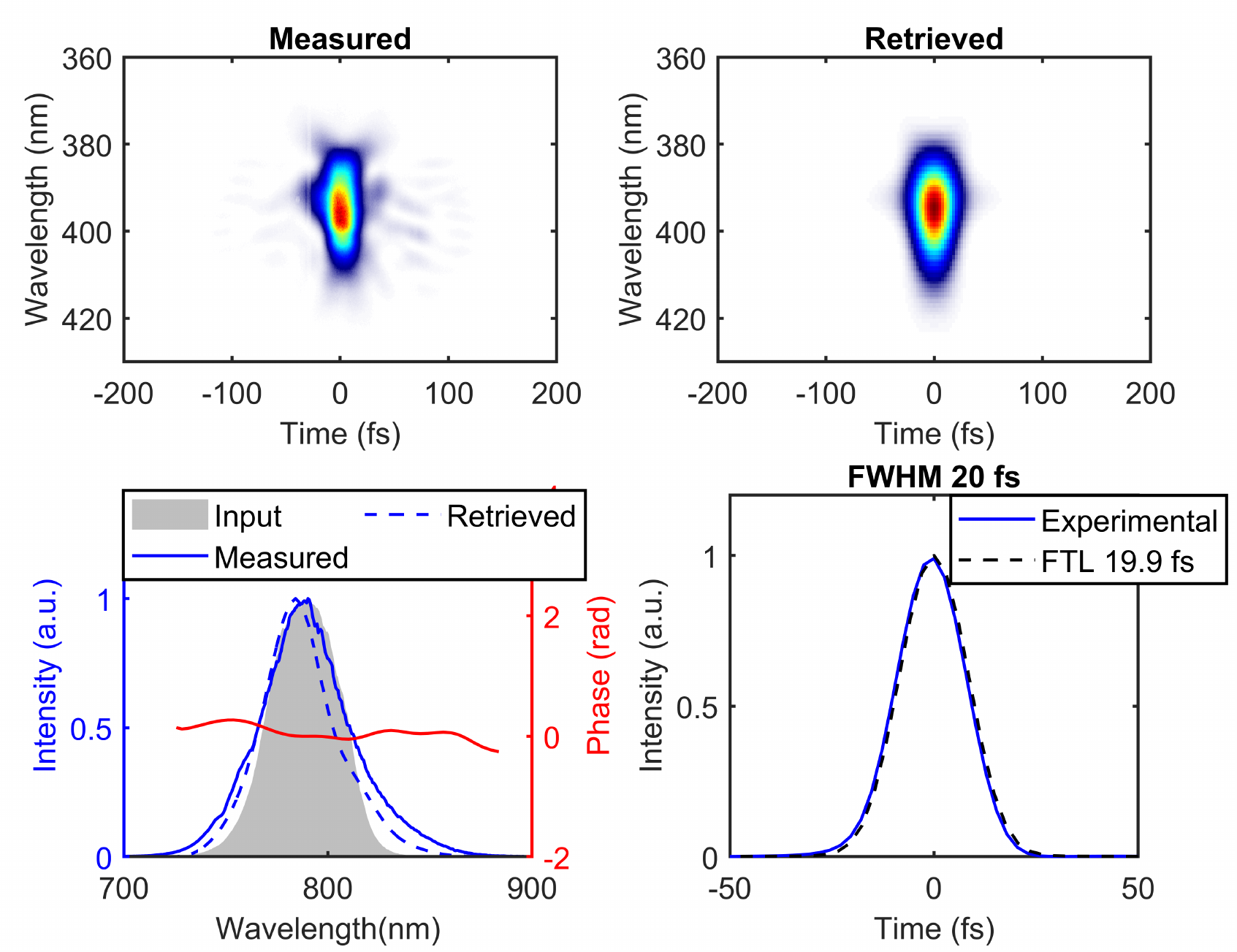}
\caption{Top: measured (left) and retrieved (right) SHG-FROG traces; bottom: input and XPW spectral properties (left) and corresponding output XPW temporal profile (right).}
\label{fig:FROG-XPW-ChirpedCase}
\end{figure}

The broadened XPW spectrum can be compressed down to 15\,fs as shown in fig. \ref{fig:FROG-XPW-FTLcase}. Fig. \ref{fig:homogeneity-XPW} shows the spectrally-resolved beam profile measured in both the vertical and horizontal dimensions with the imaging spectrometer, which exhibits a nearly-Gaussian profile in both cases. The homogeneity factor, as defined in \cite{MPC_Weitenberg2017}, is also shown for both the dimensions in fig. \ref{fig:homogeneity-XPW}. The beam exhibits excellent spectral homogeneity above 99\% at the FWHM and above 95\% at 1/e$^2$ in both dimensions. This is a direct advantage of implementing XPW in an MPC, where incremental accumulation of B-integral mitigates spatio-temporal couplings and yields excellent output beam quality. Fig. \ref{fig:contrast-XPW} shows the long-range temporal intensity profiles measured with the TUNDRA for both input and XPW pulses. The contrast enhancement is at least 3 orders of magnitude and limited, as for NER, by polarization extinction capability of the TFP. The pre-pulse at +7.5\,ps and the train of post-pulses are similar to those observed in the NER measurements.  

\subsection{Maximizing XPW efficiency}

XPW generation has been shown to be accompanied by significant spatio-temporal reshaping due to interplay between XPM and SPM involving both fundamental and XPW pulses~\cite{MoreEfficientXPW_Julien2008,XWPspatialTemporalDynamics_Adams2010}. When an intial -500~$fs^2$ spectral phase is applied to the input pulses with the Dazzler to globally compensate for the effects of dispersion inside the MPC, the XPW energy increases to $360~\mu$J, corresponding to 65\% internal and 34\% global conversion efficiencies, respectively. To our best knowledge, this is the highest conversion efficiency reported so far for XPW generation. However, this increase in conversion efficiency comes at the cost of lower spectral broadening and therefore slightly longer re-compressed pulses of 19\,fs, as shown in fig. \ref{fig:FROG-XPW-ChirpedCase}. This result is in good agreement with previous studies on the effect of residual chirp on the output spectral behaviour~\cite{FewCycle_XPW_Jullien2009}, where narrower albeit smoother output XPW spectra were observed for negatively chirped input pulses. Finally, spectral homogeneity and contrast enhancement factors similar to the FTL case were measured for negatively chirped input pulses. Overall, the smooth XPW spectrum together with the increased available XPW pulse energy could be particularly useful for efficient seeding of further laser amplification stages in a double chirped pulse amplifier architecture. 

\section{Summary}

In conclusion, we demonstrate efficient spatial-temporal cleaning of mJ-energy 30\,fs pulses in an MPC using two different third-order nonlinear filtering techniques: XPW and NER. Comprehensive (3+1)D numerical simulations show excellent agreement with the measured data in both cases and enables us to carefully design the MPC architectures so as to obtain the highest output pulse fidelity. In both cases, a contrast enhancement $>10^3$ could be observed together with near-FTL post-compressed pulse durations. 

To the best of our knowledge, this is the first time that XPW has been implemented inside an MPC, exhibiting several advantages over a more conventional free-space setup: (1) record high efficiencies (up to 65\% internal and 34\% global), (2) no need for spatial filtering, (3) excellent output beam quality and spectral homogeneity, and (4) relatively higher tolerance to input beam pointing fluctuations. More adapted surface coatings on the nonlinear crystals and cavity end-mirrors should help significantly increase the overall energy throughput and polarization optics with higher extinction ratios could easily increase the contrast enhancement factor ($>10^3$) by 2-3 orders of magnitude. This approach could therefore aid in designing efficient and compact devices for spatio-temporal pulse cleaning in high-peak-power laser systems. 

To the best of our knowledge, this is also the highest total internal efficiency (up to 69\%) reported for NER for 30\,fs input pulses. Implementing NER in an MPC architecture with such pulses enables the direct generation of high-contrast few-cycle pulses ($< \mathrm{6}\:$ fs) with up to 43\% global efficiency, in a single post-compression stage. The inherent power scalability of MPCs makes this an attractive end-of-chain solution for producing high peak-power few-cycle pulses with high temporal contrast suited to ultra-high intensity laser-matter experiments. Shorter post-compressed pulse duration, down to the single-cycle regime, could in principle be reached using dispersion-engineered coatings targeting net-zero linear chirp to suppress the saturation of Kerr nonlinearities, as observed in~\cite{SN3_MPC_Daniault2021}, which should enable octave-spanning broadening with high throughput. For this, however, the limitations on pulse compressibility imposed by the residual phase profile of these coatings and ionization remains to be investigated.

\begin{backmatter}

\bmsection{Funding} 
This work was supported by the Agence National de la Recherche (ANR-10-LABX-0039-PALM), the Horizon 2020 Framework program (Advanced Grant ExCoMet 694596), LASERLAB-EUROPE (871124) and the R\'egion Ile-de-France (SESAME 2012-ATTOLITE).

%\bmsection{Acknowledgments} 

\bmsection{Disclosures} The authors declare no conflicts of interest regarding publication of this article.

\bmsection{Data Availability Statement} Data underlying the results presented in this article are not publicly available at this time but may be obtained from the authors upon reasonable request.

% \bmsection{Supplemental document}

\end{backmatter}

\bibliography{bibList}

\end{document}